\documentclass[floatfix,aps,pra,twocolumn,a4paper,superscriptaddress,nofootinbib,balancelastpage,reprint]{revtex4-1}
\usepackage[hidelinks,breaklinks=true]{hyperref}
\usepackage[english]{babel}
\usepackage{amsopn,amsthm,dsfont}

\usepackage{mathtools}
\mathtoolsset{showonlyrefs=true}
\usepackage{tikz}
\usepackage[caption=false]{subfig}

\usetikzlibrary{positioning}

\newtheorem{theorem}{Theorem}

\newcommand{\cancel}[1]{}
\begin{document}
\title{
Non-Locality Without Counterfactual Reasoning
}
\author{Stefan Wolf}
\affiliation{Faculty of Informatics, Universit\`{a} della Svizzera italiana, Via G. Buffi 13, 6900 Lugano, Switzerland}

\begin{abstract}
	\noindent
Non-local correlations are usually
understood
through the outcomes
of alternative measurements
(on two or more parts of a system)
 that
cannot altogether actually be carried out in an experiment. 
Indeed, a joint input/output --- {\em e.g.},
measurement-setting/outcome --- 
behavior is non-local if and only if the
outputs
for  {\em all\/} possible inputs cannot coexist 
consistently. It has been argued
that this
counterfactual view
 is how  Bell's inequalities and their  violations
are to be
seen. We~propose an alternative
perspective which refrains from              setting into relation the results of
  mutually exclusive measurements,
but that is based solely on 
 data actually available.
Our approach uses {\em algorithmic complexity\/} instead of
probability,   implies non-locality to have similar consequences as in
the probabilistic view, and  is conceptually simpler yet at the
same time more general than the latter. 
\end{abstract}

\maketitle

\noindent
{\it Introduction.}---
In an eponymous text, {\em Asher Peres\/} (1978)~\cite{peres} states that ``unperformed
experiments have no results.'' He argues first that it is not only illegitimate to
speculate about unperformed experiments, but also 
that refraining 
from it frees physics of epistemological difficulties such as the EPR paradox~\cite{bell}. Indeed, already {\em Ernst
  Specker\/} (1961)~\cite{specker} had pointed out the
counterfactual nature of the reasoning against the 
possibility of embedding quantum theory into propositional logic: 
``This is  related to the scholastic
speculations on the {\em infuturabili}, {\em i.e.}, the question
whether divine omniscience  extends to {\em what would have happened
if something had happened that did not happen}.'' More recently,
Zukowski and Brukner~\cite{bruz} suggested that non-locality is to be understood in
terms of such {\em infuturabili}, called there ``counterfactual
definiteness.'' 
We
propose
an alternative line of argumentation
which
avoids relating outcomes of measurements that cannot be
mutually carried out.
Our reasoning employs {\em complexity\/} instead of
 probability:
First, 
non-locality
  can
 be 
{\em defined\/} in such a setting, and second,   it goes along with
{\em similar consequences\/} as in the probabilistic view. The new 
approach is more general but uses, at the same time,   fewer
assumptions on the conceptual level.

\noindent
{\it Non-Locality  with Counterfactual Reasoning.}---
Non-local correlations~\cite{bell} are a fascinating feature of quantum
theory. Conceptually challenging  is the difficulty~\cite{tomy},~\cite{ws} to
explain their origin {\em causally}, {\em
  i.e.},
according to {\em Reichenbach's principle}, stating that a
correlation between two space-time events
can stem from a {\em common cause\/} (in the common past) or from a {\em direct influence\/}
from one  to the other~\cite{Reichenbach}. 
More specifically, the difficulty manifests itself when {\em
  alternatives\/}
are taken into account:
The argument leading up to a Bell inequality relates outcomes of
different measurements, only {\em one\/} of which can actually be realized. Does this mean that if
we drop the assumption of {\em counterfactual
  definiteness\/}~\cite{bruz} ({\em i.e.}, of
a consistent coexistence of the consequences of mutually exclusive
starting points), then
the paradox   disappears? 
We
argue the answer to be {\em negative\/}: Even 
in the {\em factuals-only view}, the joint properties --- in terms of
mutual compressibility~--- of the
involved  pieces of information are 
remarkable since certain {\em consequences\/} of non-local
correlations,
as they are known from the  probability calculus,
persist. An example is  the complexity (instead of randomness) 
carried over from the inputs to the outputs if no-signaling holds.

In the probabilistic regime, 
a {\em Popescu-Rohrlich (PR) box\/}~\cite{pr}
gives rise to a mechanism of the following kind. Let $A$ and $B$ be the 
respective input bits to the box and~$X$ and~$Y$ the (unbiased) output
bits
with
\begin{align}
\label{prcon}
X\oplus Y=A\cdot B\, .
\end{align}
This system is {\em no-signaling}, {\em i.e.}, the joint input-output
behavior is useless for message transmission.
According to Fine~\cite{fine}, the non-locality of the
system ({\em i.e.}, conditional distribution) $P_{XY|AB}$ |  meaning 
that it cannot be written as a convex combination of products
$P^r_{X|A}\cdot P^r_{Y|B}$~|, is equivalent to the fact that there exists no
``roof distribution'' $P'_{X_0X_1Y_0Y_1}$ the marginal~$P'_{X_iY_j}$
of which equals $P_{XY|A=i,B=j}$ for all $(i,j)\in
\{0,1\}^2$:
Non-locality 
is the impossibility of 
the outputs to {\em alternative 
inputs\/} to consistently coexist. 

Such  reasoning
 assumes and concludes that 
certain pieces of {\em classical\/\footnote{{\em Classicality\/} of
information is, as  the limit of 
macroscopicity~\cite{gisinmacro} when the  representing system's size tends to infinite, 
 an
idealized notion implying that it can be measured without any
disturbance, and that the outcome
is always the same. It makes sense 
to say that a  {\em classical bit $U$  exists\/} 
({\em i.e.}, takes  a
definite value) or does not exist.}
information   exist
or  do not exist}. 
In this way of speaking, Fine's theorem~\cite{fine} reads: {\em ``The outputs
cannot  exist before the inputs do.''\/} Let us make this
qualitative statement more precise. 
We assume a perfect PR box, {\em i.e.}, a system always satisfying~\eqref{prcon}.
Note that~\eqref{prcon} alone does not uniquely
determine $P_{XY|AB}$ since the marginal of~$X$, for instance, is not
fixed. If, however, we additionally require {\em no-signaling}, 
then the marginals, such as $P_{X|A=0}$ or $P_{Y|B=0}$, must be {\em perfectly 
unbiased\/} under the assumption that all four $(X,Y)$-combinations, {\em
  i.e.},
$(0,0),(0,1),(1,0)$, and $(1,1)$, 
are possible. 
To see this, assume $P_{X|A=0,B=0}(0)>1/2$. By~\eqref{prcon}
we can conclude the same for $Y$: $P_{Y|A=0,B=0}(0)>1/2$. From
no-signaling
we get $P_{X|A=0,B=1}(0)>1/2$. Using symmetry, and no-signaling
again, 
we obtain both $P_{X|A=1,B=1}(0)>1/2$ and $P_{Y|A=1,B=1}(0)>1/2$.
This  contradicts~\eqref{prcon} since  two bits which are  both
biased towards $0$ cannot  differ with
certainty. Therefore, our original assumption was wrong: The outputs 
{\em must\/} be perfectly unbiased. Altogether, this means that~{\em $X$ as well as
$Y$  cannot  exist\/\footnote{Actually, there cannot even exist
a classical value
 {\em arbitrarily
weakly\/} correlated with either $X$ or $Y$.}
 before
$f(A,B)$ does 
for some nontrivial 
 $f\, :\, \{0,1\}^2\rightarrow \{0,1\}$}. 
The paradoxical aspect of non-locality --- at least if a causal
structure is  in   place~--- is the fact
that {\em fresh pieces of information  come to existence in a
 spacelike-separated way  that are   perfectly correlated}. 

\

\noindent
{\it Non-Locality  Without Counterfactual Reasoning.}---
We propose an understanding of non-locality without counterfactual
definiteness. 
We  use an asymptotic  {\em Kolmogorov-complexity\/} calculus 
for  binary strings to replace
the  probability calculus.

A recent article~\cite{kurz} suggests the use of
Kolmogorov complexity in the context of non-local correlations with
the objective of avoiding
 probabilities, but not of resigning from counterfactual
arguments; in fact, the outcomes of alternative measurements {\em
  are\/} assumed
to coexist in~\cite{kurz}; 
their argument   builds on the relation between these
alternative output data.

\

\noindent
{\it An asymptotic Kolmogorov calculus.}---
Let ${\cal U}$ be a fixed universal Turing machine (TM).\footnote{We
  can assume a 
{\em fixed\/}  machine here since the introduced asymptotic notions
are independent of this choice.}
For a finite
or infinite string $s$,  the {\em Kolmogorov
  complexity\/}~\cite{kol},\, \cite{text} \mbox{$K(s)=K_{\cal U}(s)$} is the length
of the shortest program for~${\cal U}$ such that the machine outputs~$s$. Note that $K(s)$ can be infinite if~$s$ is.

Let $a=(a_1,a_2,\ldots)$
be an infinite string. Then
$a_{[n]}:=(a_1,\ldots,a_n,0,\ldots)$.
We study the asymptotic behavior of, {\em e.g.}, $K(a_{[n]})\, :\, {\bf
  N}\rightarrow {\bf N}$. For this function, we simply write~$K(a)$,
similarly $K(a\, |\, b)$ for $K(a_{[n]}\, |\, b_{[n]})$,  the latter being the length of 
the shortest program outputting~$a_{[n]}$ on input~$b_{[n]}$. 
We write
\[
K(a)\approx n :\Longleftrightarrow \lim_{n\rightarrow\infty}\left(
  K(a_{[n]})/n\right) =1\ .
\]
We call a string $a$ with this property {\em  incompressible}. 
We also use $K(a_{[n]})=\Theta(n)$, as well as
\[
K(a)\approx 0 :\Longleftrightarrow \lim_{n\rightarrow\infty}\left(
  K(a_{[n]})/n\right) = 0 \Longleftrightarrow K(a_{[n]})=o(n).
\]
Note that {\em computable\/} strings $a$ satisfy $K(a)\approx 0$.

Generally, for  functions $f(n)$ and $g(n)\not\approx0$, we write 
$f\approx g$ if $f/g\rightarrow 1$.
{\em Independence of $a$ and $b$\/} is  then
\[
K(a\, |\, b)\approx K(a)
\]
or, equivalently, 
$
K(a,b)\approx K(a)+K(b).
$
If we introduce
$
I_K(x;y):=K(x)-K(x\, |\, y)\approx K(y)-K(y\, |\, x),
$
independence of $a$ and $b$ is $I_K(a,b)\approx 0$.

In the same spirit, we can define {\em conditional independence\/}: We
say that
{\em $a$ and $b$ are independent given $c$\/} if 
\[
K(a,b\, |\, c)\approx K(a\, |\, c)+K(b\, |\, c)
\]
or, equivalently, 
$
K(a\, |\, b,c)\approx K(a\, |\, c),
$
or
$
I_K(a;b\, |\, c):=K(a\, |\, c)-K(a\, |\, bc)\approx 0.
$

\

\noindent
{\it Uncomputability from PR boxes and incompressible inputs.}---
Let now $(a,b,x,y)$ be infinite  binary strings with
\begin{align}
\label{prbit}
x_i\oplus y_i= a_i\cdot b_i\mbox{\ \ for all\ \ }i=1,2,\ldots\ .
\end{align}
Although the intuition is that the strings stand for the inputs and
outputs of a PR box,  no dynamic meaning is attached
to them  (or to the ``box,'' for that matter) since there is
{\em no free choice of an input and no generation of an output in function of  this
  input\/}; all we have are four fixed strings satisfying the PR condition.
Nothing prevents us from defining this
static  situation to be {\em no-signaling\/}:
\begin{align}
\label{ns}
K(x\, |\, a)\approx K(x\, |\, ab)\mbox{\ \ \ and\ \  \ }K(y\, |\, b)\approx K(y\, |\, ab)\ .
\end{align}

\begin{figure*}
\centering
\subfloat[\label{bild1}]{
\includegraphics[scale=1]{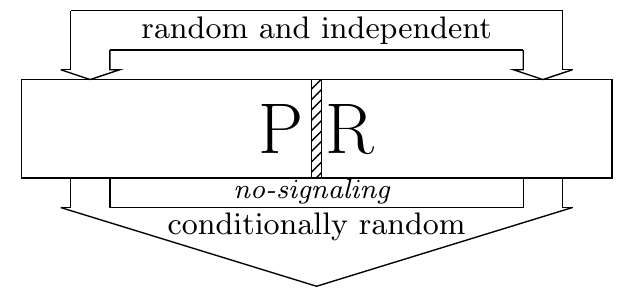}}
\qquad
\subfloat[\label{bild2}]{
\includegraphics[scale=1]{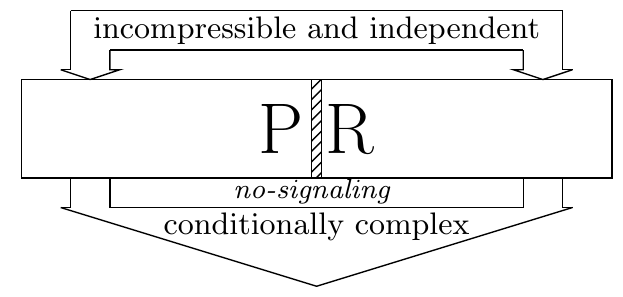}}
\caption{The traditional (a) vs.\ the new (b) view: Non-locality {\em
    \`a la\/} Popescu/Rohrlich (PR) plus
  no-signaling leads to the output inheriting {\em randomness\/} (a) or
  {\em complexity\/}~(b), respectively, from the input.}
\label{ganz}
\end{figure*}

Recall the mechanism 
enabled by
the maximal non-locality of the
PR box: {\em If the inputs are not entirely fixed, then the
outputs must be completely unbiased, as long as the system is
no-signaling.}
We can now draw a  statement of similar flavor but entirely within
{\em actual\/}
data  (see Figure~\ref{ganz}).

\begin{theorem}
If the input pair to a PR box
is  incompressible  and no-signaling holds,
then the outputs are uncomputable.
\end{theorem}

{\proof
Let $(a,b,x,y)\in (\{0,1\}^{\bf N})^4$ with~\eqref{prbit},
no-signaling~\eqref{ns}, and 
$
K(a,b)\approx 2n,
$
{\em i.e.}, the input pair is  incompressible. We  conclude 
$
K(a\cdot b\, |\, b)\approx n/2.
$
Note first  that $b_i=0$ implies $a_i\cdot b_i=0$, and second that 
any further compression of $a\cdot b$, given $b$, would lead to
``structure in~$(a,b)$,'' {\em i.e.},   a
possibility of describing or programming~$(a,b)$ in
shorter than $2n$.
Observe now
\begin{align}
\label{b0}
K(x\, |\, b)+K(y\, |\, b)\gtrsim  K(a\cdot b\, |\, b)
\end{align}
since $x$ and $y$ together determine $a\cdot b$. Now, \eqref{b0} 
implies
\begin{align}
\label{b1}
K(y\, |\, b)\gtrsim  K(a\cdot b\, |\, b)-K(x\, |\, b)\gtrsim n/2-K(x)\ .
\end{align}
On the other hand, 
\begin{align}
\label{b2}
K(y\, |\, ab)\approx K(x\, |\, ab)\leq K(x)\ .
\end{align}
Now, no-signaling~\eqref{ns}, together with~\eqref{b1} and~\eqref{b2},
implies
$
n/2-K(x)\lesssim K(x)
$,
hence,
$
K(x)\gtrsim n/4 =\Theta(n)
$.
The string $x$ must be {\em uncomputable}.
\qed}

\

Theorem~1 raises a number of  questions: Does a
similar 
result hold for the {\em conditional\/} complexities
$K(x\, |\, a)$ and~$K(y\, |\, b)$, and from {\em quantum\/} non-local 
correlations? Can we give a  {\em general definition\/} of
non-locality and does a similar result as the above hold with respect
to {\em any\/} non-local correlation? 
In the remainder, we address
these questions.

\

\noindent
{\it Conditional uncomputability of a PR box' outputs.}---
\begin{theorem}
Under the conditions of Theorem~1, we have 
$
K(x\, |\, a)=\Theta(n)\mbox{\ \ and\ \ }K(y\, |\, b)=\Theta(n)
$.
\end{theorem}

{\proof
Note
first
\[
K(x\, |\, a)\approx 0\Leftrightarrow K(x\, |\, ab)\approx  K(y\, |\, ab)\approx 0 \Leftrightarrow K(y\, |\, b)\approx
0, 
\]
{\em i.e.}, the two quantities of interest are negligible
simultaneously. 
In order to show that they are both $\Theta(n)$, we assume
 $K(x\, |\, a)\approx 0 \approx K(y\, |\, b)$ instead.
Then, there exist programs $P_n$ and $Q_n$, both of length $o(n)$, 
computing functions $f_n$ and $g_n$ with
\begin{align}
\label{fgnl}
f_n(a_{[n]})\oplus g_n(b_{[n]})=a_{[n]}\cdot b_{[n]}\ .
\end{align}
For fixed (families of) functions $f_n$ and $g_n$, asymptotically 
 how many $(a_{[n]},b_{[n]})$ can there be at most that
satisfy~\eqref{fgnl}? This question boils down to a parallel-repetition analysis of
the PR game. A result by Raz~\cite{raz} implies that this number 
is of order \mbox{$(2-\Theta(1))^{2n}$}. Therefore, the two programs $P_n$ and
$Q_n$ |  together with the index of length
$
(1-\Theta(1))2n
$
to 
 single out the correct pair $(a_{[n]},b_{[n]})$ within the candidates' list 
of length $(2-\Theta(1))^{2n}$ --- leads to a program of 
length~\mbox{$
o(n)+(1-\Theta(1))2n
$}
with output $(a_{[n]},b_{[n]})$, in conflict with $(a,b)$'s 
incompressibility. 
\qed}

\

\noindent
{\it Conditional uncomputability from  quantum
  correlations: Chained Bell inequality and magic-square game.}---
Unfortunately, the quantum-physically achievable approximations to the
PR box do not seem to allow for a similar reasoning immediately. We
can, however, use the violation of the {\em chained Bell inequality\/}
(see, {\it e.g.},~\cite{kent},\, \cite{colrenamp})
 realizable in the laboratory~\cite{tit}. 

\cancel{

In the ``traditional view'' on non-locality, the PR box is an
idealization unachievable by the behavior of any quantum state. 
If it {\em did\/} exist, on the other hand, it would be a most
precious resource, {\em e.g.}, for cryptography or randomness 
amplification. The reason is that --- as we have discussed above --- 
under the minimal assumption that the inputs are {\em not completely 
determined}, the output must be {\em perfectly random\/} (even 
given the inputs).  
The best {\em approximations\/} to PR boxes that are quantum
physically achievable ($\sim 85\%$) can be used for tasks such as 
key agreement~\cite{esth}. For our application here, however, they
seem to be too weak.

 It has been shown~\cite{kent},\, \cite{colren},\, \cite{colrenamp},
however,
that correlations which {\em are\/} achievable in the laboratory~\cite{tit}
allow for similar applications; they are based on the {\em chained
  Bell inequality\/} instead of (ideal) CHSH (PR-) non-locality.

}

To the chained Bell inequality belongs the following idealized system: 
Let $A,B\in\{1,\ldots,m\}$ be the inputs. We assume  the
``promise''
that $B$ is congruent to~$A$ or to~$A+1$ modulo~$m$. Given this, 
 the outputs $X,Y\in\{0,1\}$ must satisfy
\begin{align}
\label{chain}
X\oplus Y={\chi}_{A=m,B=1}\ ,
\end{align}
where ${\chi}_{A=m,B=1}$ is the characteristic function of the event
$\{A=m,B=1\}$. 
Barrett, Hardy, and Kent~\cite{kent} showed
that if $A$ and $B$ are random, then $X$ and $Y$ must be almost perfectly 
unbiased if the system is no-signaling. More precisely, they were 
 able to show such a statement from the gap between the 
error probability of the best quantum~($\Theta(1/m^2)$) as opposed to
classical~($\Theta(1/m)$) strategy for winning the
game. 

\begin{theorem}
Let 
$(a,b,x,y)\in (\{1,\ldots,m\}^n)^2\times (\{0,1\}^n)^2$ be 
respecting the promise
and such that
\begin{align}
\label{max}
K(a,b)\approx (\log m+1)\cdot n\ ,
\end{align}
{\it i.e.},  $(a,b)$ is  incompressible conditioned on
  the promise; the system is  no-signaling~\eqref{ns};
the fraction of quadruples $(a_i,b_i,x_i,y_i)$, $i=1,\ldots,n$, with~\eqref{chain} 
is  $1-\Theta(1/m^2)$. Then~$K(x)=\Theta(n)$. In particular, $x$ is
uncomputable.
\end{theorem}
{\proof
Since $K(a,b)$ is
maximal, we have\footnote{$h$
  is the binary entropy \mbox{$h(p)=-p\log p-(1-p)\log(1-p)$}.
Usually, $p$ is a probability, but $h$ is invoked here merely as an approximation for binomial coefficients.
}\begin{align}
\label{chib}
K(\chi_{a=m,b=1}\, |\, b)=h(\Theta(1/m))n\ . 
\end{align}
Note first that this is the maximal complexity 
of a string  in which the fraction of $1$'s is of order
$\Theta(1/m)$, since the number  of such strings is 
\[
\log {n \choose \Theta(1/m)n}\approx h(\Theta(1/m))n\ .
\]
Second, if the complexity were any lower, 
this would
lead
to the possibility of compressing $(a,b)$, which is excluded. 

\noindent
Now, we have
\[
K(x\, |\, b)+K(y\, |\, b)+h(\Theta(1/m^2))n\gtrsim K(\chi_{a=m,b=1}\, |\, b)
\]
since one possibility for generating the string $\chi_{a=m,b=1}$ 
from position $1$ to $n$ is to generate $x_{[n]}$ and $y_{[n]}$, as well as 
the string indicating the positions where~\eqref{chain}
is 
violated; the complexity of the latter is at most 
\[
\log {n \choose \Theta(1/m^2)n}\approx h(\Theta(1/m^2))n\ .
\]
Together with~\eqref{chib}, we get
\begin{align}
\label{c1}
K(y\, |\, b)\geq \Theta(1/m)n-K(x)
\end{align}
if  $m$ is sufficiently large. On the other
hand, 
\begin{align}
K(y\, |\, ab)&\lesssim K(x\, |\, ab)+h(\Theta(1/m^2))n\\
\label{c2}
&\leq K(x)+h(\Theta(1/m^2))n\ .
\end{align}

\noindent
Now,~\eqref{ns}, \eqref{c1}, and~\eqref{c2} together imply 
$
K(x)=\Theta(n).
$
\qed}

As above, the application of Raz'
parallel-repetition 
result~\cite{raz} allows for proving  
{\em conditional\/} uncomputability: 
\[
K(x\, |\, a)=\Theta(n)\ .
\]
In order to see that, note first
\[
K(y\, |\, ab)\lesssim K(x\, |\, ab)+h(\Theta(1/m^2))n\ .
\]
Therefore, the assumption
$
K(x\, |\, a)\approx K(x\, |\, ab)\approx 0
$
leads, just as above, to a program  generating $(a_{[n]},b_{[n]})$ of
length
\begin{align}
\label{rav}
&h(\Theta(1/m^2))n+(1-h(\Theta(1/m)))(\log m+1)\\
&=(1-\Theta(1))(\log m+1)\not\approx (\log m+1)n
\end{align}
for fixed, sufficiently large $m$, in conflict with the assumption
that $(a,b)$ be incompressible. 
In~\eqref{rav} we  use that for fixed output-alphabet sizes,
Raz' result bounds the success probability in the parallel repetition
of the game by $2^{-c\cdot 1/m\cdot n}$ for some constant $c$, and that \mbox{$1/m^2<c/m$} holds for 
sufficiently large $m$.

For any non-local behavior  characterizable by a 
condition that is always  satisfiable with entanglement, 
but not {\em without\/} this resource | so called ``pseudo-telepathy''
games~\cite{bbt} |, a similar reasoning  shows 
that incompressibility of the inputs leads to 
uncomputability of at least one of the two  outputs,
even given the corresponding input. We illustrate the 
argument with the example of the {\em magic-square game\/}~\cite{ara}:
Let $(a,b,x,y)\in(\{1,2,3\}^{\bf N})^2\times (\{1,2,3,4\}^{\bf N})^2$
be the quadruple of the inputs and outputs,
respectively, and assume that the pair~$(a,b)$ is incompressible
as well as $K(x\, |\, a)\approx 0\approx K(y\, |\, b)$. Then there exist
$o(n)$-length programs $P_n$, $Q_n$ such that $x_n=P_n(a_{[n]})$ and $y_n=Q_n(b_{[n]})$. 
Again, Raz' parallel-repetition theorem~\cite{raz} implies that the
length of a program generating $(a_{[n]},b_{[n]})$ is, including the employed sub-routines $P_n$ and $Q_n$,
of order 
$(1-\Theta(1))\mbox{len}(a_{[n]},b_{[n]})$ --- in violation of the incompressibility of~$(a,b)$.

\

\noindent
{\it General factual definition of (non-)locality.}---
We propose the following definition of when a no-signaling quadruple 
$(a,b,x,y)\in (\{0,1\}^{\bf N})^4$ (where $a,b$ are the inputs
and $x,y$ the outputs) is {\em local}. There  exists
$\lambda\in \{0,1\}^{\bf N}$ such that 
\begin{align}
& K(a,b,\lambda) \approx K(a,b)+K(\lambda)\ ,\label{lambdainsicht}\\
&K(x\, |\, a\lambda)  \approx 0\ , \mbox{\ and\ }
K(y\, |\, b\lambda)  \approx 0\ .
\end{align}

\noindent
Sufficient conditions for locality are 
$
K(a,b)\approx 0$ or $K(x,y)\approx 0$ 
because of $\lambda:=(x,y)$. At the other end of the scale, 
we expect that for any non-local ``system,'' the fact that $K(a,b)$ is
maximal  implies  $x$ or $y$ to be conditionally  uncomputable,
given $a$ or $b$, respectively.

It is a natural question whether the given definition harmonizes with
the 
probabilistic understanding. Indeed, the latter can be seen as a special 
case of the former: If the (fixed) strings are {\em
  typical sequences\/} of a stochastic process, our non-locality
definition
implies non-locality of the corresponding  conditional  distribution. 
The reason is that a hidden variable of the distribution 
immediately
gives rise, through sampling, to
a $\lambda$ in the sense of~\eqref{lambdainsicht}.
Note, however, 
that our formalism is more general since most strings {\em
  cannot\/} be  seen
as  typical sequences of such a  process.

\

\noindent
{\it Conclusion.}---
We  propose a  view on non-locality that does not 
rely on relating
outcomes of measurements that cannot  be actually
carried out altogether. It is based on the notion of {\em complexity\/} rather than
probability. 
In the argument, Kolmogorov complexity  --- defined with respect to Turing
machines --- can be replaced by  other computing models with finite machine
descriptions. 
While the resulting reasoning,
leading to similar
mechanisms enabled by non-locality,
is  more general than  the probabilistic regime,
a central assumption |~the \mbox{existence} of
results of unperformed measurements~---
can even be dropped.
Our statements and reasoning are asymptotic and
apply to {\em
  sufficiently long finite strings}.

In the derivation of Bell inequalities and, hence, 
the analysis of real-life experiments demonstrating non-locality, the
assumption that the outcomes of all alternative measurements  exist
together
is used to be  \mbox{implicitly} made. Our line of reasoning
 suggests a more {\em direct\/} discussion, referring only
 to
  the
data at hand. Note that whereas Kolmogorov complexity is uncomputable
itself, upper
bounds on the quantity can be obtained easily from any compression algorithm~\cite{cbc}.

\cancel{

It is natural to ask whether similar conclusions can be drawn from 
{\em contextuality\/}~\cite{specker}. 
This seems to be {\em not\/} the case
since non-contextuality translates into an independence condition
 compatible with trivial output complexities:  
Nothing seems to forbid a measured Qtrit to output the binary
expansion of $\pi$
even when some incompressible sequence determines the setting choices.
The argument of~\cite{ks}, where the contrary is claimed, 
uses the
implicit assumption that if {\em some\/} measurement results have
definite values, then {\em all\/} do. 

}

In short,
our main result is as follows. If the settings' description in a non-locality experiment is
incompressible, then the outcomes must be uncomputable, even given the respective
inputs:
{\em If the experimenter is able to generate an incompressible string,
then the measured photons must be able to come up with a non-computable 
behavior as well}. 
This gives an all-or-nothing flavor to the Church-Turing hypothesis, since
``beyond-TM'' computations either do not exist at all, or they occur even in individual
photons.
Tightened versions of our results 
give rise to a physical {\em
  incompressibility-amplification and -expansion mechanism\/}.

\noindent
{\bf Acknowledgement.}\
The author thanks Mateus Ara\'ujo, \"Amin Baumeler, Harvey Brown, Caslav Brukner, Bora Dakic, Paul
Erker, J\"urg Fr\"ohlich, Arne Hansen, Marcus
Huber, Alberto Montina, Benno Salwey, Andreas \mbox{Winter}, and Magdalena Zych for enlightening discussions. \\
The author  is supported by the Swiss National Science
        Foundation (SNF), the National Centre of Competence in
        Research ``Quantum Science and Technology'' (QSIT), and
the COST action on Fundamental Problems in Quantum Physics.
We thank all  members of the {\em Facolt\`a indipendente di Gandria\/}
for their support.



\end{document}